\documentclass[12pt,fleqn]{article}
\usepackage{graphicx}

\usepackage{latexsym}

\usepackage{amsmath}
\usepackage{amsthm}
\usepackage{amssymb}
\usepackage{amsfonts}

\begin{document}

\newcommand{\rf}[1]{(\ref{#1})}
\newcommand{\rff}[2]{(\ref{#1}\ref{#2})}

\newcommand{\ba}{\begin{array}}
\newcommand{\ea}{\end{array}}

\newcommand{\be}{\begin{equation}}
\newcommand{\ee}{\end{equation}}

\newcommand{\const}{{\rm const}}
\newcommand{\ep}{\varepsilon}
\newcommand{\Cl}{{\cal C}}
\newcommand{\rr}{{\vec r}}
\newcommand{\ph}{\varphi}
\newcommand{\R}{{\mathbb R}}  
\newcommand{\N}{{\mathbb N}}  
\newcommand{\C}{{\mathbb C}}  

\newcommand{\e}{{\bf e}}

\newcommand{\m}{\left[ \ba{c}}
\newcommand{\ema}{\ea \right]}
\newcommand{\mm}{\left[ \ba{cc}}
\newcommand{\mmm}{\left[ \ba{ccc}}
\newcommand{\miv}{\left[ \ba{cccc}}

\newcommand{\scal}[2]{\mbox{$\langle #1 \! \mid #2 \rangle $}}
\newcommand{\ods}{\par \vspace{0.5cm} \par}
\newcommand{\no}{\noindent}
\newcommand{\dis}{\displaystyle }
\newcommand{\mc}{\multicolumn}

\newenvironment{Proof}{\par \vspace{2ex} \par
\noindent \small {\it Proof:}}{\hfill $\Box$ 
\vspace{2ex} \par }

\newtheorem{prop}{Proposition}[section]
\newtheorem{Th}[prop]{Theorem}
\newtheorem{lem}[prop]{Lemma}
\newtheorem{rem}[prop]{Remark}
\newtheorem{cor}[prop]{Corollary}
\newtheorem{Def}[prop]{Definition}
\newtheorem{open}{Open problem}
\newtheorem{ex}{Example}
\newtheorem{exer}{Exercise}

\title{\bf 
Fast exact  digital differential analyzer for circle generation}
\author{
 {\bf Jan L.\ Cie\'sli\'nski}\thanks{\footnotesize
 e-mail: \tt janek\,@\,alpha.uwb.edu.pl}
\\ {\footnotesize Uniwersytet w Bia{\l}ymstoku,
Wydzia{\l} Fizyki }
\\ {\footnotesize ul.\ Lipowa 41, 15-424
Bia{\l}ystok, Poland}
\\[1ex] {\bf Leonid V.\ Moroz}\thanks{\footnotesize 
e-mail:  \tt  moroz\_lv\,@\,polynet.lviv.ua}
\\ {\footnotesize Lviv Polytechnic National University, Department of Security Information and Technology}
\\ {\footnotesize  st. Kn. Romana 1/3, 79000 Lviv, Ukraine}
}

\date{}

\maketitle

\begin{abstract}
In the first part of the paper we present a short review of applications of digital differential analyzers (DDA) to generation of circles showing that they can be treated as one-step numerical schemes. 
In the second part we present and discuss a novel fast algorithm based on a two-step numerical scheme (explicit midpoint rule).    
Although our algorithm is as cheap as the simplest one-step DDA algoritm (and can be represented in terms of shifts and additions), it  generates circles with maximal accuracy, i.e., it is exact up to round-off errors.
\end{abstract}

\ods

{\it Key words and phrases:} circle generation, digital differential analyzers, exact discretization, explicit midpoint rule

\section{Introduction}

In spite of apparent simplicity fast generation of circles and other curves has always been a subject of numerous theoretical and practical studies, finding applications, among others, in digital plotting, graphical display and numerical machine tool control, compare \cite{FGL,Ko,NS,YK}. 
Leaving aside the fundamental algoritm of  Bresenham \cite{Br} and techniques based on spline functions \cite{PT},  in this paper we focus on digital differential analyzers (DDA), see for instance \cite{Da,LWR,MB}.  
In section~\ref{sec-examples} we present brief but exhaustive discussion of digital differential analyzers from a unified point of view, treating them as special one-step difference numerical schemes.  Then, in section~\ref{sec-new}, we introduce and discuss a new DDA scheme, cheap and accurate, based on a two-step numerical scheme.

In terms of the natural arc parameter $\vartheta$ the equation of the circle of radius $r$ is $x= r \cos\vartheta$, $y=-r \sin\vartheta$. The corresponding differential equation is
\be  \label{dif-cir}
 \frac{d x}{d \vartheta} = - y \ , \quad \frac{d y}{d \vartheta} = x \ .
\ee
Considering various discretizations of \rf{dif-cir} we can obtain any DDA algorithm.

\section{One-step numerical approximations}

\label{sec-examples}

In this section  we present a unified approach to digital differential analysers representing them by one-step numerical schemes. Therefore they can be represented by a general matrix difference equation of the first order:
\be  \label{abcd}
  \m x_{n+1} \\ y_{n+1} \ema = \mm a  & b \\ c & d \ema \m x_n \\ y_n \ema , \qquad   \mm a  & b \\ c & d \ema \equiv A \ , 
\ee
where $a, b, c, d$ depend on the time step $h$.  In order to approximate the system \rf{dif-cir} these coefficients have to satisfy 
\be  \ba{l} \label{lim}   \dis
   \lim_{h \rightarrow 0} \mm a  & b \\ c & d \ema  = \mm 1  & 0 \\ 0 & 1 \ema \equiv I , \\[3ex] \dis
 \lim_{h \rightarrow 0} \frac{A-I}{h} = \mm 0  & -1 \\ 1 & 0 \ema  \equiv J  . 
\ea \ee
In other words, \ $A = I + h J + O (h^2)$, i.e.,  
\[
a = 1 + O(h^2)  , \ \   d = 1 + O(h^2)  , \ \  b = - h + O(h^2) , \ \ c = h + O(h^2) . 
\]
The qualitative asymptotic behaviour of the discrete solution $x_n, y_n$ is determined by the 
characteristic equation
\be
 0 =  \det \mm a - \lambda & b \\ c & d - \lambda \ema  \equiv \lambda^2 - (a + d) \lambda + ad- bc 
\ee
and its roots (eigenvalues of the matrix $A$) are given by
\be  \ba{l}
\lambda_1 = \frac{1}{2} \left( a + d + \sqrt{(a+d)^2 - 4 (ad - bc)} \right) =  1 + i h + O (h^2)  , \\[2ex]
\lambda_2 = \frac{1}{2} \left( a + d - \sqrt{(a+d)^2 - 4 (ad - bc)} \right)  =  1 - i h + O (h^2)   ,
\ea \ee
where $i = \sqrt{-1}$.

\subsection{Logarithmic spirals}

The most popular DDA schemes have the form 
\be  \label{ac}
  \m x_{n+1} \\ y_{n+1} \ema = \mm a &  - c  \\  c   & a  \ema \m x_n \\ y_n \ema  .
\ee
It is convenient to denote
\be
 \rho  = \sqrt{ a^2 + c^2} \ , \qquad      \theta = \arctan \frac{c}{a} \ ,
\ee
i.e., \ $a = \rho \cos\theta$, \ $c = \rho \sin\theta$. 
Then, 
\[
   \m x_n \\ y_n \ema = \mm  \rho \cos\theta & - \rho \sin\theta \\ \rho \sin\theta & \rho \cos\theta \ema^n \m x_0 \\ y_0 \ema =  \rho^n  \mm   \cos n \theta & -  \sin n \theta \\  \sin n \theta &   \cos n \theta \ema \m x_0 \\ y_0 \ema  .
\]
Denoting \ $ x_0 = r_0 \cos\ph_0$ ,  \ $y_0 = r_0 \sin\ph_0$, \  we finally get
\be
  x_{n} = r_0 \rho^n \cos (n \theta + \ph_0) \ , \quad y_n = r_0 \rho^n \sin (n \theta + \ph_0) \ . 
\ee

\begin{cor}
All points generated by \rf{ac} lie on the logarithmic spiral defined by: 
\be
  r = r_0 e^{ k (\ph - \ph_0)} \ , \qquad  k := \theta^{-1}  \ln \rho \ , 
\ee
where $r, \ph$ are polar coordinates on the plane $(x,y)$, compare \cite{BF}. 
\end{cor}

From computational point of view the polynomial form of $a, c$ is preferred. DDA algorithms can be classified by the order of  these polynomials \cite{LWR}.  Below we present several schemes of this kind with the corresponding values of $\rho^2$ and $k$ (the exact algorithm should have $\rho=1$ and $k=0$). 

\subsubsection*{First order simultaneous DDA algorithm} 
\be   \label{1dda}
a = 1 \ , \qquad  c = h \ , \qquad  \rho^2 = 1 + h^2 \ , \qquad   k = \frac{h}{2} - \frac{h^3}{12} + \ldots \ .
\ee
This is the conventional DDA method \cite{LWR}. 
 
\subsubsection*{Second order simultaneous DDA algorithm} 
\be
a = 1 - \frac{1}{2} h^2 \ , \quad c = h \ , \quad  \  \rho^2 = 1 + \frac{1}{4} h^4 \ , \quad  k = \frac{h^3}{8} - \frac{h^5}{48} + \ldots \ , 
\ee
see \cite{Be,Ko}. 

\subsubsection*{Third order simultaneous DDA algorithm} 
\be  \label{3dda}
 a = 1 - \frac{1}{2} h^2  , \quad c = h - \frac{1}{6} h^3  , \quad \rho^2 = 1 - \frac{1}{12} h^4 + \frac{1}{36} h^6 , \quad 
k = - \frac{h^3}{24} + \frac{h^5}{72} + \ldots ,
\ee
compare \cite{Be}.

\subsubsection*{Matsushiro's analyzer} 
\be  \label{Ma} 
 a = 1 - \frac{1}{2} h^2  , \quad c = h - \frac{1}{4} h^3  , \quad \rho^2 = 1 - \frac{1}{4} h^4 + \frac{1}{16} h^6 , \quad k = - \frac{h^3}{8} + \frac{h^5}{48} + \ldots . 
\ee
 An important point is that this scheme (with $h$ replaced by $\frac{1}{2} h$) was implemented in terms of shifts and additions (without using multiplications) and patented, see \cite{Mats,MO}. 

\subsubsection*{The best circular interpolator of the third order}

Interestingly enough, the Matsushiro algoritm can be easily improved by changing the last term, see \cite{Ch,Mor}:
\be  \label{Chami}
 a = 1 - \frac{1}{2} h^2  , \quad c = h - \frac{1}{8} h^3  , \quad \rho^2 = 1 + \frac{1}{64} h^6 , \quad k =  \frac{h^5}{128}   + \ldots . 
\ee
We are going to show that this is the best spiral-like circular interpolator of the third order in $h$. 
\ods

\begin{lem}  \label{lem-Chami}
The most accurate circular interpolator of the form \rf{ac}, where $a, c$ are polynomials of the third order in $h$, is given by
\rf{Chami}
\end{lem}

\begin{Proof} \ We assume \ 
\be 
a = 1 + a_1 h + a_2 h^2 + a_3 h^3 \ , \quad  c = h + c_2 h^2 + c_3 h^3
\ee
Then  we compute $\rho^2 = a^2 + c^2$:
\[
\rho^2  = 1 + 2 a_1 h + (2 a_2 + a_1^2 + 1) h^2 + 2 (a_3 +  a_1 a_2 +  c_2) h^3 + \ldots + (a_3^2 + c_3^2) h^6 . 
\]
Equating to zero five first coefficients we obtain a system of 5 equations for 5 unknowns, which can be solved easily:
\be  \ba{l}  \label{5eqs}
2 a_1 = 0 \  \quad \Longrightarrow \quad   a_1 = 0 \ , \\[1ex]
2 a_2 + a_1^2 + 1 = 0 \  \quad \Longrightarrow \quad  a_2 = - \frac{1}{2} \ ,    \\[1ex]
a_3 + a_1 a_2 + c_2 = 0 \  \quad \Longrightarrow \quad  a_3 = - c_2 \ ,  \\[1ex]
a_2^2 + 2 a_1 a_3 + c_2^2 + 2 c_3 = 0 \  \quad \Longrightarrow \quad  c_3 = - \frac{1}{8} - \frac{1}{2} c_2^2 \ ,   \\[1ex]
a_2 a_3 + c_2 c_3 = 0  \   \quad \Longrightarrow \quad  \frac{1}{2} c_2  \left( \frac{3}{4} - c_2^2 \right) = 0 \ .
\ea  \ee
Then, the coefficient by $h^6$ is given by \ $a_3^2 + c_3^2 = c_2^2 +  \frac{1}{4}\left( \frac{1}{4} + c_2^2 \right)^2$. The last equation of \rf{5eqs} yields  two possibilities: either $c_2 =0 $ or $c_2^2 = \frac{3}{4}$. Corresponding coefficients by $h^6$ are given by $\frac{1}{64}$ and $1$, respectively. Therefore the best accuracy is attained in the first case, which leads to \rf{Chami}  (the second case has another disadvantage: coefficients $c_2$ and $a_3$ are irrational). \par 
Interestingly enough, considering $h$-expansion of the function $k=k(h)$ we get the same final results (although the starting system of  equations is more complicated).  The case  $c_2 =0$  yields $k = \frac{1}{128} h^5 + O (h^6)$ while in the second case, $c_2^2 = \frac{3}{4}$, we get $k = \frac{1}{2} h^5 + O (h^6)$. 
\end{Proof}

\subsection{Elliptical deformations}

Discrete points generated by \rf{abcd} lie on an ellipse  if and only if  \ $\bar \lambda_2 = \lambda_1$ and  $|\lambda_1|=1$, which is equivalent to   
\be
|a + d| < 2 \quad \text{and} \quad  a d - b c =1 \ .
\ee
The conditions are known as Barkhausen criteria \cite{Tu}.

\subsubsection*{First order sequential DDA algorithm}

\be
  \m x_{n+1} \\ y_{n+1} \ema = \mm 1  & - h  \\ h  & 1 - h^2  \ema \m x_n \\ y_n \ema .
\ee
This is the generator of the so called ``magic circle'' \cite{Mats,NS,Tu}.

\subsection{Elliptical spirals} 

Two complex eigenvalues  ($\lambda_2 = {\bar \lambda}_1$) are obtained when 
\be  \label{elspir}
(a-d)^2 + 4 bc < 0 \ , 
\ee
and, for $a d - bc \neq 1$  the generated circle deforms into an elliptical spiral. The following example can be found in \cite{LWR}. 

\subsubsection*{Second order sequential DDA algorithm} 

\be \label{2-seq}
  \m x_{n+1} \\ y_{n+1} \ema = \mm 1 - \frac{1}{2} h^2  & - h  \\  h  & 1 - \frac{3}{2} h^2  \ema \m x_n \\ y_n \ema 
\ee
One can easily verify, that $(a-d)^2 + 4 b c = h^2 (h^2 -4)$. Therefore \rf{elspir} is satisfied for $|h|<2$. 

Sequentials algorithms are derived from simulataneous algorithms by replacing $x_n$ by $x_{n+1}$ in the equation for $y_{n+1}$, i.e., simultaneous algorithm \rf{ac} of  order $M$ is transformed into
\be \ba{l} 
x_{n+1} = a x_n - c y_n \ , \\
y_{n+1} = c x_{n+1} + a y_n = a c x_n + (a - c^2) y_n \ ,
\ea \ee
and then truncated up to the order $M$, see \cite{LWR}. In the case $M=2$ we get the scheme \rf{2-seq}.

\subsection{Exactly circular interpolation} 

Digital differential analyzer of the form \rf{abcd} yields an exact circle (up to round-off errors) if and only if 
$a = d$, $c=-b$ and $a d - bc =1$. Then, the matrix $A$ can be parameterized by a single parameter $\theta$
\be
     A = \mm \cos\phi & - \sin\phi \\ \sin\phi & \cos\phi \ema ,
\ee  
where $\phi = \phi (h)$ depends on time step $h$. In other words, 
\be  \label{CS}
  \m x_{n+1} \\ y_{n+1} \ema = \mm C (h)  & - S (h)  \\ S (h)  & C (h)  \ema \m x_n \\ y_n \ema 
\ee
where $C^2 (h)  + S^2 (h) = 1$. 
Taking $\theta (h) = h$ we obtain the exact numerical scheme (compare \cite{NS,WW}) 
\be  \label{1-exact}
  \m x_{n+1} \\ y_{n+1} \ema = \mm \cos h  & - \sin h  \\ \sin h  & \cos h   \ema \m x_n \\ y_n \ema .
\ee 
Exact numerical schemes can be constructed for any matrix differential equation of the first order, see \cite{CR-ade,Ci-oscyl}.

The Taylor expansion of \rf{1-exact} up to the third order yields the scheme \rf{3dda}, compare \cite{Be}. We recall that this is not the best scheme of the third order, see  Lemma~\ref{lem-Chami}.  

\subsubsection*{Implicit midpoint rule} 

In the case of linear equations the implicit midpoint rule can be represented in an explicit form. Indeed,  
\be  
   \frac{x_{n+1} - x_n}{h} = - \frac{y_n + y_{n+1}}{2} \ , \qquad 
\frac{y_{n+1} - y_n}{h} = - \frac{x_n + x_{n+1}}{2} \ , 
\ee
can be rewritten in a matrix form as follows
\be
  \mm 1  & \frac{1}{2} h \\ - \frac{1}{2} h  & 1 \ema \m x_{n+1} \\ y_{n+1} \ema = 
\mm 1  & - \frac{1}{2} h \\  \frac{1}{2} h  & 1 \ema \m x_n \\ y_n \ema .
\ee
Thus we obtain  another algortihm of the form \rf{ac}, where
\be   \label{ac-imp}
   a = \frac{4  - h^2}{4 + h^2} \ , \qquad c = \frac{4 h}{4 + h^2} \ .
\ee
see \cite{Ko}. 
Note that $a^2+c^2 = 1$ and $k=0$, so the implicit midpoint rule generates the circle exactly. However, this algorithm has a relatively high cost (multiplications in every step). Actually the scheme  \rf{ac-imp} was a motivation for the derivation of 
the Matsushiro algorithm \rf{Ma}, see \cite{Mats}. Indeed, \rf{Ma} is the Taylor truncation of \rf{ac-imp} up to the third order.

\section{New fast  circular interpolator}

\label{sec-new}

The main result of our paper consists in applying a two step method to the circular interpolation. 
The explicit midpoint rule (see \cite{Is}), known also as 2-step Nystr\"om method \cite{HNW}, applied to a general ordinary differential equation  \  $\pmb{\dot z} = \pmb{f} (t, \pmb{z})$ \ (where $\pmb{z} \in \R^n$ and $\pmb{f}$ is a given function) reads: 
\be  \label{emp}
\pmb{z}_{n+2} = \pmb{z}_n + 2 h \pmb{f} (t_{n+1}, \pmb{z}_{n+1}) \ .
\ee
In the case of the circular interpolation  \ $\pmb{z} = \m x \\ y \ema$ and $\pmb{f} (t, \pmb{z}) \equiv \m - y \\ x \ema$. Therefore we obtain: 
\be \ba{l}  \label{ex-mid}
x_{n+2} = x_n - 2 h y_{n+1} \ , \\[1ex]
y_{n+2} = y_n + 2 h x_{n+1} \ , 
\ea \ee
compare \cite{Mor}. This is a two step method and we have to prescribe not only  $x_0, y_0$ but also \  $x_1, y_1$. As usual, we can rewrite this scheme in a one-step form increasing the dimension of the matrix problem, namely:
\be   \label{emr-matrix} 
\m x_{n+2} \\ y_{n+2} \\ x_{n+1} \\ y_{n+1} \ema = \miv  0 & - 2 h & 1 & 0 \\ 2 h & 0 & 0 & 1 \\ 1 & 0 & 0 & 0 \\ 0 & 1 & 0 & 0 \ema  \m x_{n+1} \\ y_{n+1} \\ x_n \\ y_n \ema .
\ee

\subsection{First integrals}

In this section we are going to show that  \rf{ex-mid} preserves circular trajectories exactly provided that initial conditions are appropriately chosen. 

\begin{lem}
If \ $x_n^2 + y_n^2$ \ is a first integral of the system \rf{ex-mid}, then \  $x_n y_{n+1} - x_{n+1} y_n$ \ is also a first integral. What is more, in this case both first integrals are linearly dependent, namely 
\be
   x_n y_{n+1} - x_{n+1} y_n = h (x_n^2 + y_n^2)
\ee
\end{lem}

\begin{Proof}  \  Using \rf{ex-mid} we compute
\[
x_{n+2}^2 + y_{n+2}^2 = x_n^2 + y_n^2 - 4 h (x_n y_{n+1} - x_{n+1} y_n) + 4 h^2 (x_{n+1}^2 + y_{n+1}^2) \ .
\]
If \ $x_n^2 + y_n^2 = r^2 = \const$, then, obviously,  \ $x_n y_{n+1} - x_{n+1} y_n = h r^2$. 
\end{Proof}

Motivated by above results, we define: 
\be   \label{xiB}
\pmb{\xi}_n = \m x_n^2 + y_n^2 \\ x_{n+1}^2 + y_{n+1}^2 \\ x_n y_{n+1} - x_{n+1}  y_n  \ema , \quad B = \mmm  0 & 1 & 0 \\ 1 & 4 h^2 & - 4 h \\ 0 &  2 h &  - 1 \ema .
\ee
\ods
\begin{lem}   For any $x_n, y_n$ satisfying \rf{ex-mid} we have
\be  \label{xin}
  \pmb{\xi}_{n+1} = B \pmb{\xi}_n , \qquad 
\ee
\end{lem}
\begin{Proof} \ A simple straightforward calculation.  \end{Proof}

\begin{Th}  \label{Th-circle}
All points generated by  \rf{ex-mid}  lie exactly (up to round-off errors) on the circle $x + y = r^2$ provided that initial conditions satisfy 
\be  \label{init}
x_0^2+y_0^2= r^2  , \qquad  x_1^2+y_1^2 = r^2  , \qquad    x_0 y_1 - x_1 y_0 = h r^2  . 
\ee 
\end{Th}

\begin{Proof}  \  We easily verify that 
\be  \label{f1}
    B f_1 = f_1 \ , \qquad  \pmb{f}_1 := \m 1 \\ 1 \\ h \ema .
\ee
Initial conditions \rf{init}  can be shortly written as  $\pmb{\xi}_0 = r^2 \pmb{f}_1$. By \rf{xin} we have  $\pmb{\xi}_n =\pmb{\xi}_0$  for any $n$. It means that \ $x_n^2+y_n^2 = r^2$ \ for any $n$.  \end{Proof}

\begin{rem} The simplest initial conditions satisfying \rf{init} read
\be \label{incon}
   x_0 = r \ , \quad  y_0 = 0 \ , \quad  x_1 = r\, \sqrt{1-h^2} \ , \quad  y_1 = h r \  ,
\ee
compare \cite{Mor}.  
\end{rem}

\subsection{Low computational cost}

The scheme \rf{ex-mid} has a very low computation cost, to be compared only with the simplest one-step scheme \rf{1dda}. Indeed, both \rf{ex-mid} and \rf{1dda} can be implemented in terms of two additions and two shifts. However,   \rf{ex-mid} produces an exact circle while the accuracy of \rf{1dda} is rather limited. 

There is only one more complicated computation, at the start, when initial conditions are $x_1, y_1$ are determined. However, taking into account  the Taylor expansion 
\be
\sqrt{1-h^2} = 1 - \frac{1}{2} h^2 - \frac{1}{8} h^4 - \frac{1}{16} h^6 - \frac{5}{128} h^8 - \frac{7}{256} h^{10} + O(h^{12})
\ee
we see that even in that case one can use few shifts to produce $x_1$. Indeed, let $h=2^{-m}$ and $r =2^N$. Then
\be
   \sqrt{1-h^2} = 1 - 2^{-2m -1} - 2^{-4 m - 3} - 2^{- 6 m - 4} - 2^{- 8 m - 5} - 2^{- 8 m - 7} + \ldots  
\ee
and 
\be
  x_1 = 2^N  - 2^{N -2m -1} - 2^{N -4 m - 3} - 2^{N - 6 m - 4} - 2^{N - 8 m - 5} - 2^{N - 8 m - 7} + \ldots  
\ee
where only integer coefficients are left. 

The case $h=\frac{1}{2}$ (i.e., $m=1$) corresponds to the  circle approximated by regular dodecagon (because then $x_1 =   r \cos \frac{\pi}{6}$, $y_1 = r \sin \frac{\pi}{6}$, compare \rf{incon}). Assuming $N=8$ we may confine ourselves to the first three terms.

\subsection{Stability}

Characteristic equation $\det (B - \lambda I) = 0$ reads
\be
  (1-\lambda) (\lambda^2 + 2 (1 - 2 h^2) \lambda + 1) = 0 ,
\ee
hence eigenvalues of $B$  are given by: 
\be 
\lambda_1 = 1 ,  \ \   \lambda_2 = 2 h^2 - 1 + 2 h \sqrt{h^2-1} , \ \ \lambda_3 = 2 h^2 - 1 - 2 h \sqrt{h^2-1}  . 
\ee
All the eigenvalues lie on the unit circle: $|\lambda_k|=1$ for $k=1,2,3$. Corresponding eigenvectors are denoted by $\pmb{f}_k$, $k=1,2,3$, namely
\be  \label{Bf}
  B \pmb{f}_1 = f_1 \ , \quad B \pmb{f}_2 = \lambda_2 \pmb{f}_2 \ , \quad B \pmb{f}_3 = \lambda_3 \pmb{f}_3 ,
\ee
($\pmb{f}_1$ is given by \rf{f1} and the form of other eigenvectors is not important). 
The initial condition $\pmb{\xi}_0$ can always be represented as a linear combination of eigenvectors: $\pmb{\xi}_0 = a \pmb{f}_1 + b \pmb{f}_2 + c \pmb{f}_3$, where we assume $a \approx r^2$ (compare Theorem~\ref{Th-circle}) and $b$, $c$ are small perturbations. 
Using \rf{Bf} and \rf{xin} we obtain
\be
\pmb{\xi}_{n} =  \pmb{f}_1 + b   \lambda_2^n \pmb{f}_2 + c \lambda_3^n \pmb{f}_3 .
\ee
Taking into account \ $|\lambda_2^n| = |\lambda_3^n| = 1$, we see that the initial perturbations remain unchanged during the evolution.

\subsection{Exact circular interpolator}

The scheme \rf{ex-mid} produces exact circular trajectory but the period is modified. 
For instance, in the case of regular dodecagon ($h = \frac{1}{2}$) the total change of $\vartheta$ after 12 steps is $\Delta \vartheta = 12 h = 6$ while the exact value is, of course, $2 \pi$. In general, the 
 period of the circular interpolator \rf{ex-mid} is given by 
\be
     T = \frac{2\pi h}{\arcsin h} .
\ee

\ods
In order to preserve exactly  not only the trajectory but also the period of the circular motion, we can use a nonstandard modification of  \rf{ex-mid}, following the approach of \cite{CR-ade,Ci-oscyl}: 
\be \ba{l}  \label{ex-mid-delta}
x_{n+2} = x_n - 2 \delta   y_{n+1} \ , \\[1ex]
y_{n+2} = y_n + 2 \delta   x_{n+1} \ , 
\ea \ee
where $\delta = \delta (h)$ is a given function of $h$.  

\begin{Th}  \label{Th-circle-del}
For any $\delta=\delta(h)$  all points generated by  \rf{ex-mid-delta}  lie exactly (up to round-off errors) on the circle $x + y = r^2$ provided that initial conditions satisfy 
\be  \label{init}
x_0^2+y_0^2= r^2  , \qquad  x_1^2+y_1^2 = r^2  , \qquad    x_0 y_1 - x_1 y_0 = \delta  r^2  . 
\ee 
\end{Th}
\begin{Proof}
It is enough to repeat the proof of Theorem~\ref{Th-circle} (with $h$ replaced by $\delta$ in appropriate places).  
\end{Proof}

The period of the circular interpolator \rf{ex-mid-delta} is 
\be
 T = \frac{2 \pi h}{\arcsin \delta (h)} .
\ee
\ods
\begin{Th}
 \label{Th-exact}
If  \ $\delta = \sin h $, then scheme 
\be \ba{l}  \label{ex-mid-sin}
x_{n+2} = x_n - 2  y_{n+1} \sin h  \ , \\[1ex]
y_{n+2} = y_n + 2  x_{n+1} \sin h  \ , \\[1ex]
x_0 y_1 - x_1 y_0 =    r^2 \sin h  \ ,
\ea \ee
generates the circle exactly (up to round-off errors) preserving the period.   
\end{Th}

\begin{Proof}
The scheme \rf{ex-mid-delta} will preserve the period exactly if \  $x_n = r \cos h n$ and  $y_n = r \sin h n$. Taking into account well known trigonometric identities
\be \ba{l}
\cos (h n + 2 h) - \cos (h n) = - 2 \sin h \sin (h n + h) \ , \\[1ex]
\sin (h n + 2 h) - \sin (h n) =  2 \sin h \cos (h n + h) \ , 
\ea \ee
we conclude that scheme \rf{ex-mid-delta} is exact for $\delta (h) = \sin h$. 
\end{Proof}

The exact interpolator \rf{ex-mid-sin} uses two multiplication by $\sin h$ at every step, hence it is relatively expensive (but two times cheaper than its one-step conterpart \rf{1-exact}).  

The scheme \rf{ex-mid-delta} can serve as a starting point for deriving cheaper  algorithms by taking $h$-polynomials instead of  $\sin h$.  All these schemes preserve exactly the circular trajectory. The best approximation of the period are obtained for Taylor truncations. In particular, taking 
\be
  \delta = h - \frac{1}{6} h^3 
\ee
we get the best scheme \rf{ex-mid-delta} of order 3.  The multiplication by $\frac{1}{6}$ can be replaced by shifts because
\be
  \frac{1}{6} =  \frac{1}{8 \left( 1 - \frac{1}{4} \right) } = \frac{1}{8} \sum_{k=0}^\infty 4^{-k} = \sum_{k=0}^\infty 2^{- 3 - 2 k} ,
\ee
and we can approximate $\delta$ by $\delta_N$ given by
\be
 \delta_N = h - h^3 \sum_{k=0}^N 2^{-3-2 k} ,
\ee
where $N$ is chosen to assure the required accuracy.

\section{Conclusions}

We discussed DDA circular interpolators based on one-step numerical schemes. We have shown that the best cheap interpolator of the third order is given by \rf{Chami}, see Lemma~\ref{lem-Chami}. This algorithm can be expressed in terms of shifts and additions only, and it has quite good accuracy. Exact interpolators based on one-step methods are more expensive because multiplications in every step are required. 

In section~\ref{sec-new} we presented a family of   circular interpolators, based on a two-step method, namely  the explicit midpoint rule. All of them interpolate the circle exactly (up to round-off errors). The simplest interpolator, given by \rf{ex-mid}, is very cheap and is very good for graphical applications (when the period of the circular motion is irrelevant). The scheme \rf{ex-mid-sin} preserves exactly also the period but is more expensive. Taking an appropriate polynomial  $\delta(h)$ we can control the accuracy and computational cost of  the scheme  \rf{ex-mid-delta}, obtaining  interpolators suitable for different  purposes.

\ods
{\it Acknowledgement.} The first author (J.L.C.) is partly supported  by the National Science Centre (NCN) grant no. 2011/01/B/ST1/05137.

\end{document}